\documentclass{ws-procs9x6-cpt16}
\begin{document}

\newcommand{\refeq}[1]{(\ref{#1})}
\def\etal {{\it et al.}}

\title{Results and Prospects from the \\ Daya Bay Reactor Neutrino Experiment}

\author{A.\ Higuera}

\address{Department of Physics, University of Houston, Houston, Texas 77204, USA}

\author{On behalf of the Daya Bay Collaboration}

\begin{abstract}
The Daya Bay reactor experiment  
has reported the most precise measurement of sin$^{2}2\theta_{13}$ and $\Delta m^{2}_{ee}$ by using a dataset with the fully constructed design of 8 antineutrino detectors. We also report on a new independent measurement of sin$^{2}2\theta_{13}$ from neutron capture on hydrogen, which confirms the results using  gadolinium captures. Several other analyses are also performed, including a measurement on the absolute reactor antineutrino flux and a search for light sterile neutrinos. Prospects for new analyses such as searching for CPT- and Lorentz-invariance violation at Daya Bay are ongoing.   

\end{abstract}

\bodymatter

\section{The Daya Bay reactor neutrino experiment}
The Daya Bay experiment is located near the Daya Bay nuclear power plant, with six reactors, two at Daya Bay and four at Ling Ao. These reactors provide a total of 17.6 GW thermal power, producing one of the most intense reactor antineutrino fluxes available in the world. The Daya Bay experiment was designed to provide the most precise measurement of $\theta_{13}$, and this has been achieved by a careful design of antineutrino detectors (ADs) and baseline. Prior to August 2012 Daya Bay had six functionally identical ADs, and after August 2012 a total of eight functionally identical ADs were installed. Each AD was filled with a 20-ton gadolinium-doped liquid scintillating (Gd-LS) target region. The ADs were distributed among near and far experimental halls. The two near halls contained four ADs (two at each experimental hall) and the far hall contained four ADs. The ADs were submerged in a water pool, with at least 2.5 m of high-purity water in all directions to shield against ambient radiation and for background tagging. Full information about the Daya Bay AD system and site can be found in Ref.\ \refcite{DayaBayNIM}.

Antineutrinos are detected via the inverse beta decay (IBD) process, $\bar{\nu}_{e} + p \rightarrow e^{+} +n$. To select IBD candidates a coincidence of the prompt signal $e^{+}$ and the delayed signal from neutron capture on Gd is required. To achieve precise measurements characterization of detector response of each AD is essential. Regular calibration runs are done using sources deployed  by automated calibration units  ($^{68}$Ge, $^{60}$Co, $^{241}$Am$^{13}$C). The difference in reconstructed energy between ADs is less than 0.2$\%$ in the energy range of the antineutrino flux. In addition characterization of the prompt signal requires precision on the absolute energy scale for $e^{+}$, $e^{-}$, and $\gamma$. The absolute energy scale uncertainty, which is correlated among all ADs, is constrained to be less than 1$\%$ in the majority of the energy range of the antineutrino flux. More information on calibration and absolute energy scale can be found in Ref.\ \refcite{DayaBayNIM}.
In the following section we discuss the analyses presented at the Seventh Meeting on CPT and Lorentz Symmetry (CPT16).

\section{Latest Daya Bay results}

\begin{figure}
\begin{center}
\includegraphics[width=2.1in,height=1.3in]{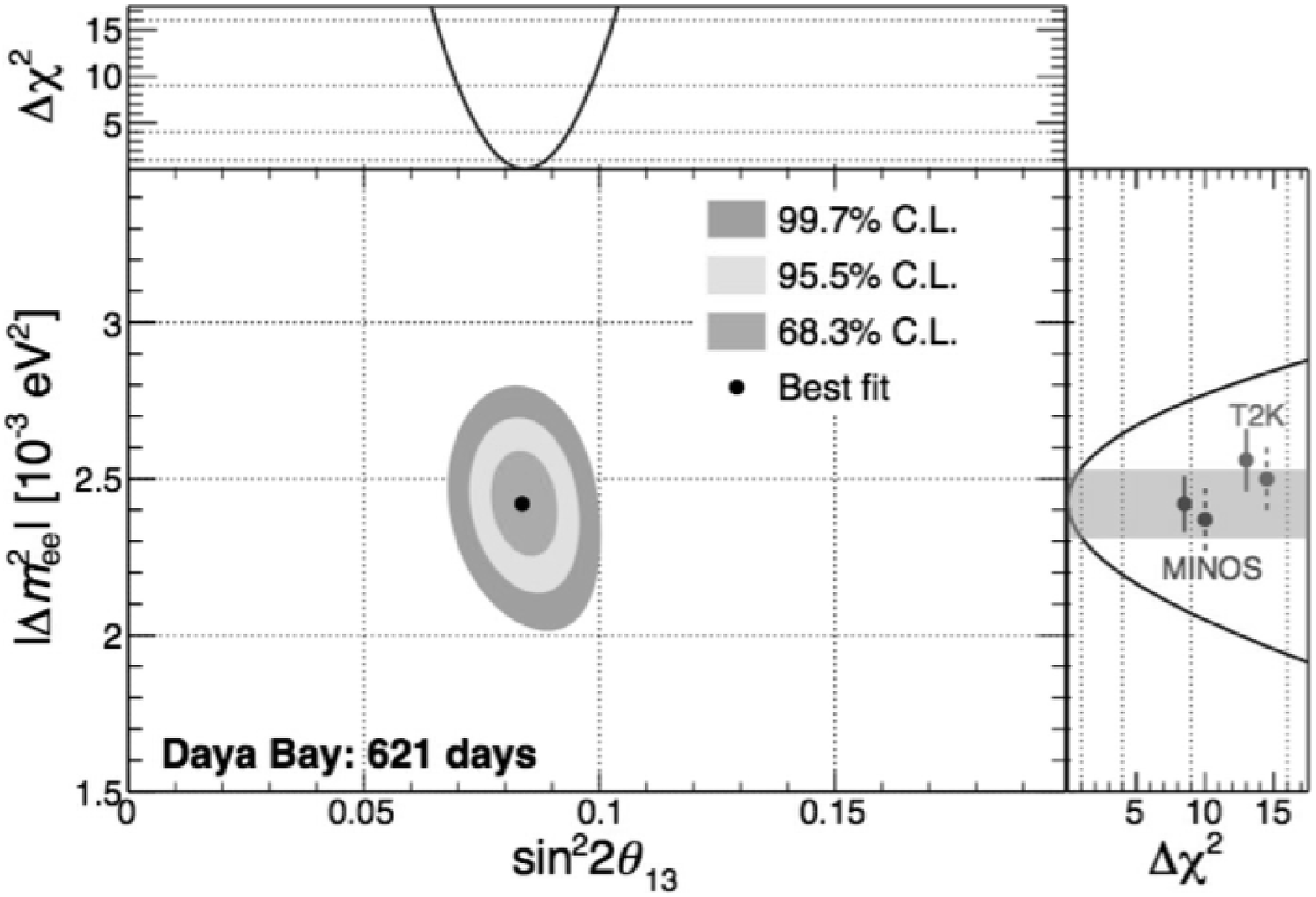}
\hskip 10pt
\includegraphics[width=2.1in,height=1.3in]{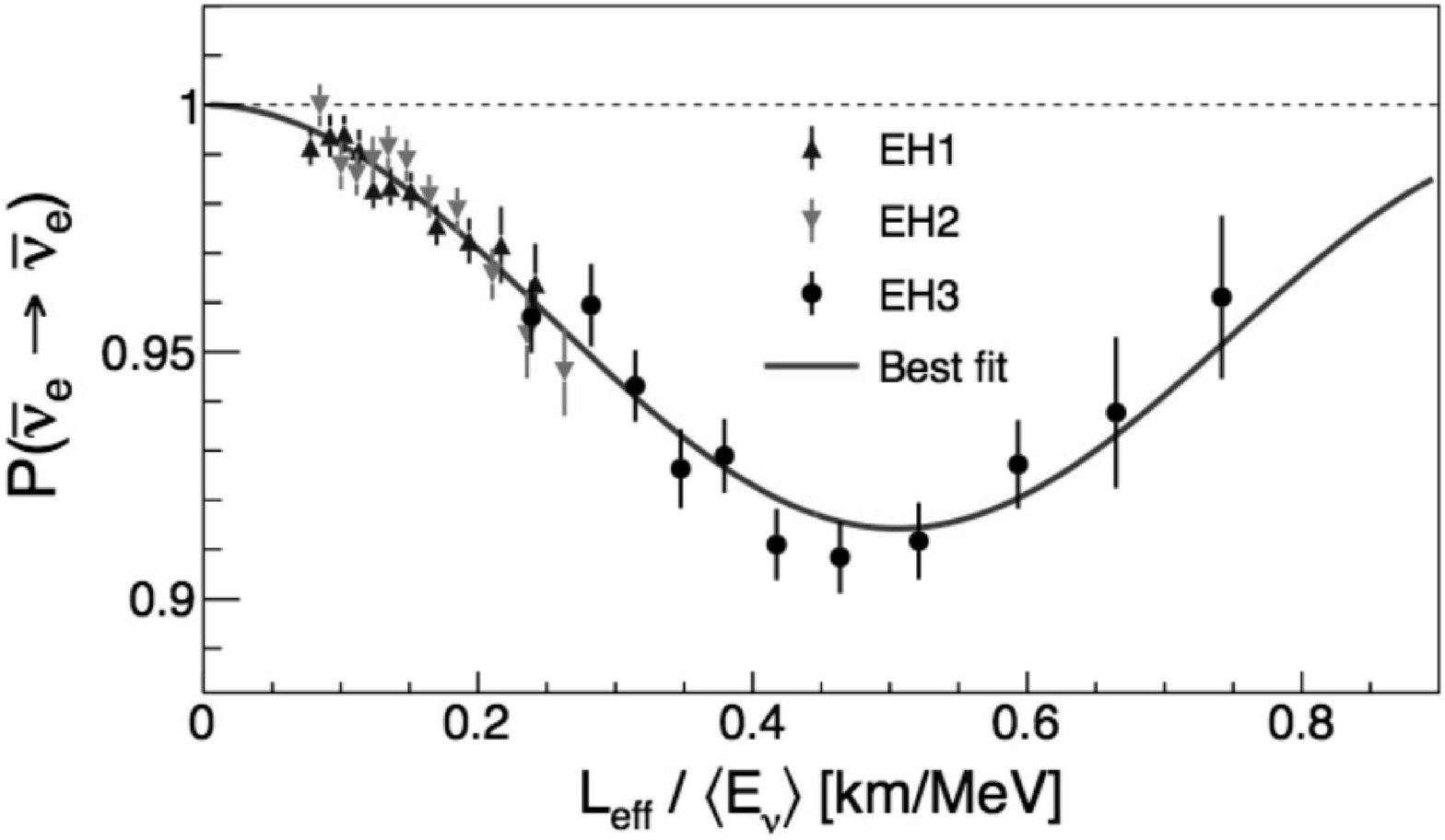}
\end{center}
\caption{Regions in $|\Delta m^{2}_{ee}|-$sin$^{2}2\theta_{13}$ plane allowed at 68.3$\%$, 95.5$\%$ and 99.7$\%$ confidence levels by the near-far comparison of the $\bar{\nu}_{e}$ rate and energy spectra (left).  Electron antineutrino survival probability versus effective propagation distance $L_{\rm eff}$ divided by the average antineutrino energy (right).}
\label{fig:Osc}
\end{figure}

During December 2011 until November 2013 a total of 1.1 million (150k) IBD candidates were selected in the near (far) halls, representing the largest  sample among all previous and current reactor neutrino experiments. The relative uncertainty in the efficiency between ADs is 0.2$\%$. The largest source of background is accidental coincidence of singles events, which accounts for 2.3$\%$ (1.4$\%$) of the candidates in the far (near) hall. Other sources of background are cosmogenic backgrounds, beta delayed-neutron emitters $^{9}$Li and $^{8}$He, fast neutrons produced by untagged muons, and background due to $(\alpha$, n) nuclear reaction. All together these are less than 0.5$\%$ of the candidates. The final estimation of signal and background rates, as well as the efficiencies of the muon veto and multiplicity selection are summarized in Table I of Ref.\ \refcite{DayaBay8ADPRL}. 
To estimate the oscillation parameters, the survival probability was measured using the L/E-dependent disappearance of $\bar{\nu}_{e}$,
\begin{equation}
P = 1 - \mbox{cos}^{4}\theta_{13}\ \mbox{sin}^{2}2\theta_{12}\ \mbox{sin}^{2}\frac{1.267 \Delta m^{2}_{21} L}{E} - \mbox{sin}^{2}2\theta_{13}\  \mbox{sin}^{2}\frac{1.267 \Delta m^{2}_{ee} L}{E},
\end{equation}
where $E$ is the energy in MeV of the $\bar{\nu}_{e}$, $L$ is the distance in meters from the $\bar{\nu}_{e}$ production point, $\theta_{12}$ is the solar mixing angle, and $\Delta m^{2}_{21}= m^{2}_{2}-m^{2}_{1}$ is the mass-squared difference of the first two neutrino mass eigenstates in eV$^{2}$. With increased statistics, improvements in calibration, and improved background estimation, we were able to provide the most precise measurement to date of $|\Delta m^{2}_{ee}|$ and sin$^{2}2\theta_{13}$. The best estimates were sin$^{2}2\theta_{13}= 0.084 \pm 0.005$ and $|\Delta m^{2}_{ee}|=(2.42 \pm 0.11)\times 10^{-3}$ eV$^{2}$.  Fig.~\ref{fig:Osc} shows regions in the $|\Delta m^{2}_{ee}|-$sin$^{2}2\theta_{13}$ plane (left) and the electron antineutrino survival probability (right). Full details of the oscillation analysis can be found in Ref.\ \refcite{DayaBay8ADPRL}. 

An independent measurement of sin$^{2}2\theta_{13}$ was reported using IBD with the emitted neutron captured by hydrogen. The deficit in the detected number of antineutrinos at the far detectors relative to the expected number on the near detectors yielded sin$^{2}2\theta_{13}$ = 0.071 $\pm$ 0.011 in the three-neutrino-oscillation framework. This result is consistent with the Gd measurement. Full details of the hydrogen analysis can be found in Ref.\ \refcite{DayaBayPRD}.

Thanks to the large $\bar{\nu}_{e}$ sample collected it is also possible to perform a precise measurement of the absolute reactor antineutrino flux. This analysis uses the 217-day dataset of the 6-AD period. The ratio of the Daya Bay measurement to the Huber-Mueller (Ref.\ \refcite{Huber,Muller}) model prediction is R = 0.947 $\pm$ 0.022, while R = 0.992 $\pm$ 0.023 when compared to the ILL-Vogel (Ref.\ \refcite{ILL,Vogel}) model prediction. These results are consistent with previous short-baseline results. Full details of this analysis can be found in Ref.\ \refcite{DayaBayReactor}.

A search for light sterile neutrino mixing was performed using the 6-AD period. The relative spectral distortion due to the disappearance of $\bar{\nu}_{e}$ was found to be consistent with that of the three-flavor oscillation model. Thus no evidence of sterile neutrinos was found. The derived limits on sin$^{2}2\theta_{14}$ cover the $10^{-3}$ eV$^{2}$ $\lesssim$ $|\Delta m^{2}_{41}|$   $\lesssim $ 0.1 eV$^{2}$ region, which was largely unexplored. Full details of this analysis can be found in Ref.\ \refcite{DayaBaySterile}.

\section{Perspectives for searching for CPT and Lorentz violation}
A physical system is said to have Lorentz symmetry if the relevant laws of physics are unaffected by Lorentz transformations. This yields a natural connection with CPT in the sense that certain theories (local quantum field theories) with Lorentz symmetry must also have CPT symmetry like the Standard Model (SM). A generalization of the SM and General Relativity that has all the conventional properties, but allows for Lorentz-invariance (LI) and CPT violation is called the Standard-Model Extension (SME); see for instance Ref.\ \refcite{SME}. The SME provides a quantitative description of Lorentz and CPT violation controlled by a set of coefficients whose values are to be determined or constrained by experiments. One of the consequences of introducing CPT/LI violation is directional dependence in an event rate as function of  sidereal time in a Sun-centered system. This can be translated into new oscillation effects, 
\begin{equation}
\begin{split}
P_{\bar {\nu}_{e} \rightarrow \bar{\nu}_{e} }= P^{(0)}_{\bar {\nu}_{e} \rightarrow \bar{\nu}_{e} } + 2L \ \mbox{Im}({S}^{(0)*}_{ee} M^{(1)}_{ee})\ \{ (\mathcal C)_{cd} + ( \mathcal A_{s})_{cd}\  \mbox{sin} \omega_{\otimes}T_{\otimes} \\ +  (\mathcal A_{c})_{cd} \ \mbox{cos} \omega_{\otimes}T_{\otimes} +(\mathcal B_{s})_{cd} \ \mbox{sin} 2 \omega_{\otimes}T_{\otimes} + (\mathcal B_{c})_{cd} \  \mbox{cos} 2 \omega_{\otimes}T_{\otimes} \},
\end{split}
\end{equation}
where $P^{(0)}$ is the SM oscillation probability and the sidereal amplitudes $(\mathcal A_{s}),\ (\mathcal A_{c}), \ (\mathcal B_{s}),\  (\mathcal B_{c})$ depend on the directional factors and the CPT/LI violation coefficients. 
See Ref.\ \refcite{Jorge} for more information.
Thanks to the Daya Bay design we are able to look at multiple beam directions from multiple baselines, and we can derive limits on individual CPT/LI violation coefficients.  With high statistics (621 days) and reduced systematic errors we expect to produce a significantly improved result for CPT/LI violation coefficients using antineutrinos from reactors.

\end{document}